\documentclass[12pt,aps]{revtex4}
\usepackage{amsmath,amssymb,graphicx}
\begin{document}

\title{Thermodynamics and area in Minkowski space:\\
Heat capacity of entanglement.}
\author{Ram Brustein, Amos Yarom }
\address{ Department of Physics, Ben-Gurion University,
Beer-Sheva 84105, Israel \\
{\rm E-mail:} {\tt ramyb@bgumail.bgu.ac.il,
yarom@bgumail.bgu.ac.il } }

\begin{abstract}

Tracing over the degrees of freedom inside (or outside) a
sub-volume $V$ of Minkowski space in a given quantum state
$|\psi\rangle$, results in a statistical ensemble described by a
density matrix $\rho$. This enables one to relate quantum
fluctuations in $V$ when in the state $|\psi\rangle$, to
statistical fluctuations in the ensemble described by $\rho$.
These fluctuations scale linearly with the surface area of $V$. If
$V$ is half of space, then $\rho$ is the density matrix of a
canonical ensemble in Rindler space. This enables us to `derive'
area scaling of thermodynamic quantities in Rindler space from
area scaling of quantum fluctuations in half of Minkowski space.
When considering shapes other than half of Minkowski space, even
though area scaling persists, $\rho$ does not have an
interpretation as a density matrix of a canonical ensemble in a
curved, or geometrically non-trivial, background.

\end{abstract}

\maketitle
\section{Introduction}
\label{S:introduction}

The microscopic origin of Black hole (BH) thermodynamics, and in
particular area scaling of black-hole entropy and other
thermodynamical quantities is still not well understood even many
years after its discovery. There have been many attempts to
explain BH thermodynamics. One of the ideas underlying several
proposals is that the horizon prevents observers from having
access to the whole set of quantum states, and hence the entropy
they measure is due to entanglement.

In an attempt to understand BH entropy Srednicki \cite{Srednicki},
(and previously Bombelli et. al. \cite{Bombelli}) considered the
von Neumann entropy of quantum fields in a state described by the
density matrix $\rho_{in}$, defined by taking the vacuum state and
tracing over the degrees of freedom (DOF) external to a spherical
sub-volume of Minkowski space. Thus, effectively reducing the DOF
available to an observer, to the inside of the spherical
sub-volume. They discovered that this ``entropy of entanglement"
was proportional to the surface area of the sphere. Using general
arguments it is possible to show that the entropy obtained by
tracing over the DOF outside of a sub-volume of any shape is equal
to the one obtained by tracing over the DOF inside this
sub-volume. However, a numerical calculation was needed  in order
to show that it is linearly dependent on the area (for a spherical
sub-volume). Following this line of thought, some other geometries
were considered, in particular, half of Minkowski space in various
dimensionalities (see for example,
\cite{Holzheyetal,CalWil,SusUgl,deAlwis1,deAlwis2}).

We offer a novel point of view and a new calculational tool to
establish area scaling properties of statistical averages in such
a setup. Recently, we have shown \cite{HDR} that the variance of a
class of operators restricted to the interior/exterior of a
sub-volume $V$ of Minkowski space scale linearly with the boundary
area of the sub-volume. Specifically, we have considered
operators, $O$, in d+1 dimensional Minkowski space, which are
obtained by integrating over an operator density $\mathcal{O}$:
$O=\int \mathcal{O} d^dx$. One can restrict such operators to the
inside of a sub-volume $V$ by defining $O^{V} =\int_V \mathcal{O}
d^dx$. Under certain conditions, fluctuations of $O^{V}$ in the
vacuum state $\langle 0|(O^V-\langle O^V\rangle)^2|0\rangle$,
scale linearly with the boundary area of $V$. In this setup, the
vacuum state, being a superposition of external products of states
in the `in' and `out' region, is an entangled state.

To show how area scaling of fluctuations of operators in the
vacuum of Minkowski space are related to area dependence of
thermodynamic quantities, we recall that expectation values of
operators restricted to a certain region of space may be described
by an average over a mixed state described by a density matrix
$\rho_{in}$ \cite{Feynmanstatmech} (see eq.~(\ref{VN})). One may
then derive area scaling of statistical averages in the system
described by $\rho_{in}$, from area scaling properties of quantum
fluctuations in the Minkowski vacuum \cite{HDR}. Since
fluctuations of operators are related to thermodynamic quantities,
we can establish that the thermodynamics described by $\rho_{in}$
scales linearly with the surface area of $V$. Equation (\ref{VN})
and the properties of $\rho_{in}$ are reviewed in section
\ref{S:transitiontothermal}.

As an explicit example of this relation, we consider the case
where the volume in question is half of space. This was studied
earlier by \cite{CalWil,KabStr,Holzheyetal}, where it was shown
that $\rho_{in}=\langle e^{-\beta_{0} H_{R}} \rangle $, $H_{R}$
and $\beta_{0}$ being the Rindler Hamiltonian and inverse
temperature. Equation (\ref{VN}) allows us to relate the energy
fluctuations in Rindler space to fluctuations of the Rindler
Hamiltonian, in the Minkowski vacuum. As is well known, energy
fluctuations are a measure of the heat capacity. Therefore, we
have here another method by which one can show that the heat
capacity of Rindler space scales linearly with the surface area of
the horizon. This is done in section
\ref{S:heatcapacityforrindler}.

Obviously, one would like to know if tracing over DOF inside (or
outside) sub-volumes of general shape would correspond to other
spaces, with properties that are, perhaps, similar to Rindler
space. That is, we may always define an effective Hamiltonian
$\langle e^{-\beta_0 H_{eff}}\rangle=\rho_{in}$, and ask if
$H_{eff}$ describes a system in a static curved background. We
show that for shapes other than half of space,  it is not. This is
done in section \ref{S:notforothershapes}. We conclude with a
summary and discussion of our results.

\section{Properties of the density matrix $\rho_{in}$.}
\label{S:transitiontothermal}

The density matrix describing the Minkowski vacuum state is
$\rho=|0\rangle\langle0|$. Now define
$\rho_{in}=\text{Trace}_{out}\rho$. Where `$\text{Trace}_{out}$'
means tracing over those states which are external to the
sub-volume $V$. The density matrix obtained this way has many
interesting properties \cite{CalWil,KabStr, Srednicki,
Feynmanstatmech, deAlwis1,Holzheyetal}. We review some of them.

First, we wish to recall the correspondence between quantum
expectation values of operators $O^V$ restricted to a sub-volume
$V$, and statistical averages in a system described by a density
matrix $\rho_{in}$. This is a well known feature of quantum
mechanics, and it is essential for our discussion. Writing $
    |0\rangle=\sum A_{\alpha,a}|a\rangle \otimes |\alpha\rangle,
$ where $|a\rangle$ is a state inside the volume $V$ and
$|\alpha\rangle$ is external to it,  it follows that
$
    \rho_{in}=\sum_{\gamma} \langle\gamma|0\rangle\langle0|\gamma\rangle=\sum A_{\alpha,a} |a\rangle
        \langle b| A_{b,\alpha}^{\dagger}.
$ One may now show, by expressing $O^V$ in the `in' basis, and
performing some straightforward algebra \cite{Feynmanstatmech},
that
\begin{equation}
    \langle0|O^V|0\rangle=\text{Tr} (\rho_{in} O^V),
    \label{VN}
\end{equation}
so that expectation values of such operators in the vacuum of
Minkowski space are equal to expectation values of operators in
the system described by the density matrix $\rho_{in}$.

Next, we wish to find an effective Hamiltonian $H_{eff}$ which is
proportional to the logarithm of $\rho_{in}$. This can be done
using an extension of the method introduced in \cite{KabStr}: we
shall write out both $\rho_{in}$ and $\langle
e^{-\beta_{0}H_{eff}}\rangle$ as a path integral, and compare the
expressions.

Switching to a Euclidian metric, it is possible to represent the
wave functional of the vacuum as
$
      \langle 0|\psi(\vec{x})\rangle=
      \int_{\varphi(\vec{x},0)=\psi(\vec{x})}
            \exp\left[-\int_0^{\infty}\ldots\int\mathcal{L} d^dx d\tau\right]
      D\varphi,
$
where the integral is over all fields evaluated at $t > 0$, which
satisfy the boundary conditions
$\varphi(\vec{x},0)=\psi(\vec{x})$. In the specific case of
interest,
\[
    \langle0|\psi_{in}\psi_{out}\rangle=\int_
        {\varphi(\vec{x},0)=
            \begin{cases}
                    {\scriptstyle x \in V,} & {\scriptstyle \psi_{in}(\vec{x})} \\
                    {\scriptstyle x \notin V,} & {\scriptstyle \psi_{out}(\vec{x})} \\
            \end{cases}}
        \exp\left[-\int_0^{\infty}\ldots\int\mathcal{L} d^dx d\tau\right]
        D\varphi.
\]
It follows that
\begin{equation}
\label{E:rhoin}
    \langle\psi_{in}^{\prime}|\rho_{in}|\psi_{in}^{\prime\prime}\rangle
    =\int_
        {\varphi(\vec{x},0)=
            \begin{cases}
                    {\scriptstyle x \in V, t=0^{+},} & {\scriptstyle \psi_{in}^{\prime}(\vec{x})} \\
                    {\scriptstyle x \in V, t=0^{-},} & {\scriptstyle \psi_{in}^{\prime\prime}(\vec{x})}
            \end{cases}}
        \exp\left[-\int_{-\infty}^{\infty}\ldots\int\mathcal{L} d^dx d\tau\right]
        D\varphi.
\end{equation}
The effective Hamiltonian $H_{eff}$, is found by comparing
expression (\ref{E:rhoin}) to that for matrix elements of
$e^{-\beta_0 H_{eff}}$ \cite{deAlwis1,deAlwis2}. To obtain these,
we observe that
\begin{multline*}
    \langle\psi_{in}^{\prime}|e^{-\beta_0 H_{eff}}|\psi_{in}^{\prime\prime}\rangle\\
        =\int_{\substack{\varphi(x,0)=\psi_{in}^{\prime}(\vec{x}) \\
            \varphi(x,-i\beta_0)=\psi_{in}^{\prime\prime}(\vec{x})}}
        \exp\left[
            \imath\int_0^{\imath\beta_0} \left(
                \int \pi \frac{d\varphi}{dt} d^dx - H_{eff} \right) dt
            \right]
        D\varphi D\pi.
\end{multline*}
This equality is obtained by time slicing the interval
$[0,\beta_0]$ into infinitesimal values, and using the fact that
$e^{-\beta_0 H_{eff}}$ is the generator of time translations for
imaginary time, from $t=0$ to $t=-\imath \beta_0$. One must assume
here that $H_{eff}$ is time independent. Note that $\beta_0 \neq
0$.

Changing the time variable to $\tau=i t$ , $\langle e^{-\beta_0
H_{eff}}\rangle$ reduces to:
\begin{equation}
\label{E:ebheffgeneralform}
    \langle e^{-\beta_0 H_{eff}} \rangle
    =\int_{\substack{\varphi(x,0)=\psi_{in}^{\prime}(\vec{x}) \\
            \varphi(x,\beta_0)=\psi_{in}^{\prime\prime}(\vec{x})}}
        \exp\left[
            \int_0^{\beta_0} \left(
                \int \imath \pi \dot{\varphi} d^dx - H_{eff} \right) (d\tau)
            \right]
        D\varphi D\pi.
\end{equation}
If we are to compare equation (\ref{E:ebheffgeneralform}) with
equation (\ref{E:rhoin}) we must make some sort of guess about the
nature of $H_{eff}$. We therefore assume that $H_{eff}$ is the
Hamiltonian of a free field theory in a static, curved background.
Since $H_{eff}$ is the generator of time translations, it is also
the Legendre transform of a Lagrangian. A free field Lagrangian
density in a gravitational background with metric $g_{\mu\nu}$ is
given by $
    \mathcal{L}=-\frac{1}{2}\sqrt{-g}g^{\mu\nu}
        \partial_\mu\varphi \partial_\nu\varphi
$. In order to emphasize that this is not necessarily a Euclidian
parametrization, we write the space-time coordinates as
$(\vec{\xi},\eta)$. Considering static coordinates, for which
$g^{0i}=0$, and $\frac{d}{dx^0}g^{\mu\nu}=0$, then gives us
$H_{eff}=\int\mathcal{H}\sqrt{h}d^dx$, where $
    \mathcal{H}=\sqrt{-g_{00}}
    \left(\frac{1}{2}\left(\frac{\pi}{\sqrt{h}}\right)^2+\frac{1}{2}h^{ij}\partial_i\varphi\partial_j\varphi\right),
$
$h^{ij}$ are the
space-space components of the metric ($h^{ij}=g^{ij}$), and
$\pi=\frac{\dot{\varphi}\sqrt{h}}{\sqrt{-g_{00}}}$.

Performing the Gaussian integral and using
$\Omega=\frac{1}{\sqrt{g_{00}^E}}$, (where we have used
$g_{00}^{E}=-g_{00}$), one obtains the following expression
\cite{deAlwis1},
\begin{equation}
\label{E:rhothermal}
    \langle\psi_{in}^{\prime}|e^{-\beta_0 H_{eff}}|\psi_{in}^{\prime\prime}\rangle=
    \int_{\substack{\varphi(\xi,0)=\psi_{in}^{\prime}(\xi) \\
            \varphi(\xi,\beta_0)=\psi_{in}^{\prime\prime}(\xi)}}
        \exp\left[
            -\int_0^{\beta_0} \int
                \mathcal{L}_{E}
            d^d \xi d\eta
            \right]
        \left|\Omega\right|\,\left|g\right|^{\frac{1}{4}}
        D\varphi.
\end{equation}

If we find a metric under which equations (\ref{E:rhoin}), and
(\ref{E:rhothermal}) are equal, then $H_{eff}$ would be the
Hamiltonian of a free field theory in a static background given by
$g^{\mu\nu}$. In order for such an equality to be satisfied we
need: (1) That the boundary conditions be identical. (2) That the
actions be identical. (3) That the measures be identical. In order
to satisfy the three conditions, we wish to find a coordinate
transformation from a Minkowski $(\vec{x},t)$ system, to a
$(\vec{\xi},\eta)$ system. In this coordinate system we
immediately get:
\begin{enumerate}
    \item
        The boundary conditions on the path integral are met if
        \begin{subequations}
        \label{E:bc}
        \begin{align}
            \label{E:bc1}
            \{(\vec{x},0^{+})|\vec{x} \in V\}&=\{(\vec{\xi},0)|\vec{\xi} \in I_\xi\}\\
            \label{E:bc2}
            \{(\vec{x},0^{-})|\vec{x} \in V\}&=\{(\vec{\xi},\beta_0)|\vec{\xi} \in I_\xi\}.
        \end{align}
        \end{subequations}
    \item
        The actions will be equal only if (a) $\mathcal{L}$ is the Lagrangian density of
        a free field theory in a Minkowski space background. So $(\vec{\xi},\eta)$ constitute
        a coordinate transformation of Minkowski space. (b) The limits of
        integration on the Lagrangian densities are equal. This requires that the transformation
        $t \to t(\vec{\xi},\eta),\vec{x} \to(\vec{\xi},\eta)$, be onto, such that the
        range of $\eta$ is $[0,\beta_0]$.
    \item
        There is an incompatibility of the measure
        \cite{deAlwis1,deAlwis2}, which we may ignore, since for
        our purposes it is a field independent multiplicative
        factor in the path integral.
\end{enumerate}

If $V$ is half of space, then as explicitly shown in
\cite{KabStr}, the ($\vec\xi,\eta$) coordinate system is a polar
coordinate system in the (x-t) plane of Minkowski space. Such a
coordinate system describes Rindler space. The effective
Hamiltonian, $H_{eff}$, is the Rindler Hamiltonian $H_R$, and
$\beta_0=\frac{2\pi}{a}$, where $a$ is a dimensional constant
introduced so that $\eta$ will have units of length. Combining
this observation with equation (\ref{VN}), we see that expectation
values of fluctuations of operators in half of Minkowski space are
equal to expectation values of the same operators in Rindler space
at some temperature $\beta_0$,
\begin{equation}
\label{E:relate_Min_to_Rin}
    \langle (\Delta O^{V_{1/2}})^2 \rangle =
    \text{Trace} (\rho_{in} (\Delta O^{V_{1/2}})^2) =
    \text{Trace} (e^{-\beta_0 H_R} (\Delta O^{V_{1/2}})^2).
\end{equation}

Looking at eq.(\ref{E:relate_Min_to_Rin}) from right to left,
then, since Rindler space thermodynamics is known to be
area-dependent \cite{deAlwis1,deAlwis2,emparan}, we should not be
surprised that fluctuations of these operators scale linearly with
the surface area of the horizon. An alternate viewpoint is that we
have ``derived" area-thermodynamcs of Rindler space from ``known"
area scaling of vacuum expectation values in Minkowski space.

\section{The heat capacity of Rindler space radiation.}
\label{S:heatcapacityforrindler}

As a concrete example of the usage of equation
(\ref{E:relate_Min_to_Rin}), we calculate the heat capacity of
Rindler space radiation. Generally, the heat capacity is
proportional to energy fluctuations: $\langle(\Delta E
)^{2}\rangle = C_{V}T^{2}$. Therefore, the area dependence of the
heat capacity of Rindler space is given by the vacuum expectation
value of fluctuations of the Rindler Hamiltonian (equation
(\ref{VN})),
\begin{equation}
\label{E:CV_Min_to_Rin}
    C_V T^2 =
    \text{Trace} (e^{-\beta_0 H_R} (\Delta H_R)^2) =
    \langle (\Delta H_R)^2 \rangle.
\end{equation}
This, in turn, has area scaling properties due to the general
argument given in \cite{HDR,tbp}, and as shown explicitly below.

We shall use the d+1 dimensional Rindler metric in the form:
\begin{equation}
\label{E:rindlerspacemetric}
    ds^2=-e^{2a\xi}d\eta^2+e^{2a\xi}d\xi^2+{{d\vec{x}_{\bot}}}^2
\end{equation}
that is obtained from the Minkowski space metric:
\[
    ds^2=-dt^2+dz^2+{{d\vec{x}_{\bot}}}^2
\]
by a coordinate transformation:
\begin{align*}
    t(\xi,\eta)&=\frac{1}{a}e^{a\xi}\sinh a\eta\\
    z(\xi,\eta)&=\frac{1}{a}e^{a\xi}\cosh a\eta\\
    \vec{x}_{\bot}&=\vec{x}_{\bot}.
\end{align*}
Here $\vec{x}_{\bot}$ stands for the $d-1$ transverse coordinates.

To find the energy fluctuations of Rindler space, we first find an
expression for the Rindler space Hamiltonian:
\begin{align*}
    H_R&=\int\sqrt{h}\sqrt{-g_{00}}\left[
            \frac{1}{2}\left(\frac{\pi}{\sqrt{h}}\right)^2+\frac{1}{2}h^{ij}\partial_i\varphi\partial_j\varphi\right]
        d\xi d \vec{x}_{\bot}\\
    &=\int e^{a\xi} e^{a\xi}\left[
            \frac{1}{2}\left(
                \partial_{\eta}\varphi\right)^2 e^{-2a\xi}
            +\frac{1}{2}\left(
                (\partial_{\xi}\varphi)^2e^{-2a\xi}+(\partial_{{\vec{x}}_{\bot}}\varphi)^2\right)
            \right]
        d\xi d \vec{x}_{\bot}\\
    &=\frac{1}{2}\int
        \left[(\partial_{\eta}\varphi)^2
        +(\partial_{\xi}\varphi)^2+e^{2a\xi}(\partial_{{\vec{x}}_{\bot}}\varphi)^2\right]
    d\xi d\vec{x}_{\bot}.
\end{align*}
To get acquainted with this operator we compare it to the
Minkowski space energy operator in half space, given by:
\[
    E_{M+}=\frac{1}{2}\int_{z>0}
        \left[
            (\partial_t\varphi)^2+(\partial_z\varphi)^2+(\partial_{{\vec{x}}_{\bot}}\varphi)^2
        \right]
    dz d\vec{x}_{\bot},
\]
where $z$ is the coordinate along which space is divided into two
halves ($z>0$ and $z<0$). In order to relate $H_R$ and $E_{M+}$  we
note that:
\[
    \left(
        \begin{array}{c}
            \partial_t \\
            \partial_z \\
        \end{array}
    \right)
    =
    e^{-a\xi}
    \left(%
        \begin{array}{cc}
            \cosh a\eta & -\sinh a \eta \\
            -\sinh a \eta & \cosh a \eta \\
        \end{array}%
    \right)
    \left(
        \begin{array}{c}
            \partial_\eta \\
            \partial_\xi \\
        \end{array}
    \right),
\]
therefore, evaluating $E_{M+}$ at an equal time contour, we have
\begin{equation}
\label{E:hmtohr}
    E_{M+}=\frac{1}{2}\int e^{-a\xi}\left[
        (\partial_\eta \varphi)^2+
        (\partial_\xi \varphi)^2+
        e^{2a\xi}(\partial_{{\vec{x}}_{\bot}})^2\right]
        d\xi d\vec{x}_{\bot}.
\end{equation}

On the other hand, evaluating $H_R$ at $\eta=0$, we have
\begin{align}
\notag
    H_R&=\frac{1}{2}\int e^{a\xi} \left[
        (\pi)^2+
        (\partial_z\varphi)^2+
        (\partial_{{\vec{x}}_{\bot}})^2\right]
        dz d\vec{x}_{\bot}.\\
\label{E:hrtohm}
    &=\frac{1}{2}\int (az) \left[
        (\pi)^2+
        (\partial_z\varphi)^2+
        (\partial_{{\vec{x}}_{\bot}})^2\right]
        dz d\vec{x}_{\bot}.
\end{align}
From equation (\ref{E:hmtohr}), we see that the energy density in
half of Minkowski space is equal to the Rindler energy density
restricted to a region close to the Rindler horizon ($\xi \to
-\infty $).

Now that we have an expression for the Hamiltonian of Rindler
space in terms of the coordinates of Minkowski space, we may
evaluate the energy fluctuations in Rindler space by using
equation (\ref{VN}),
\begin{equation}
\langle0_M|(:H_R:)^2|0_M\rangle=\text{Trace}\left[e^{-\beta_0
H_R}(:H_R:)^2\right].
\end{equation}

We carry out this calculation in the appendix. The result is
\begin{equation}
    \langle(:H_R:)^2\rangle= V_{\bot} a^2 \Lambda^{d-1}
              \frac{(d+1) \Gamma^2\left(\frac{d+1}{2}\right)}
                   {(d-1)2^{d+5} \pi^{1+\frac{d}{2}}
                    \Gamma(2+\frac{d}{2})},
\end{equation}
where $\Lambda$ is the UV cutoff of the theory, $V_\bot$ is the
area of the boundary between the two halves of Minkowski space and
$\frac{a}{2\pi}$ is the Rindler temperature. The heat capacity is
then given by:
\begin{equation}
C=\frac{\langle\Delta E^2\rangle}{T^2} =
\frac{V_\bot}{\Lambda^{2}} \frac{(d+1)
\Gamma^2\left(\frac{d+1}{2}\right)}
                   {(d-1)^2 2^{d+5} \pi^{1+\frac{d}{2}}
                    \Gamma(2+\frac{d}{2})},
\end{equation}
which scales linearly with the surface area of half space.

We summarize what we have learned so far: area scaling is a
general property of fluctuations of a certain class of operators
restricted to a sub-volume of Minkowski space. Area dependence of
the heat capacity of Rindler space radiation, is a special case of
this general property, when the volume is chosen to be half of
Minkowski space.

\section{$H_{eff}$ for other shapes.}
\label{S:notforothershapes}

Equation (\ref{VN}) suggests that the area scaling of the
entangled system in Minkowski space \cite{tbp} and the
thermodynamics of the space given by the Hamiltonian $H_{eff}$ are
two aspects of the same phenomenon. As shown explicitly in the
previous section, this allows one to obtain, for example, area
scaling of the heat capacity of a free field in Rindler space from
area scaling of quantum fluctuations in half of Minkowski space.

One would wish to show a similar equivalence for sub-volumes of
other shapes. However, as we will show, for volumes other than
half space, the effective Hamiltonian obtained cannot be that of a
non interacting field theory in a static curved, or geometrically
non-trivial, background.

Let us consider a general sub-volume $V$. Equating equations
(\ref{E:rhoin}) and (\ref{E:rhothermal}), we wish to find an onto
coordinate transformation
\begin{subequations}
\label{E:coortans}
\begin{align}
    t&=t(\eta, \vec{\xi})\\
    \vec{x}&=\vec{x}(\eta, \vec{\xi}),
\end{align}
\end{subequations}
such that $\eta\in[0,\beta_0]$, and $\vec{\xi} \in I_\xi$, and
$\frac{d}{dt}g^{\mu\nu}=0$ and $g^{0i}=0$. We note that for such a
coordinate system time ($\eta$) translations are an isometry, so
that $\frac{\partial}{\partial \eta}=K^{\mu}=(1,0,\ldots,0)$ is a
Killing vector \footnote{
    Actually, this statement is correct even if $g^{0i}\neq 0$. However,
    we will see that once we impose the boundary conditions, we shall get $g^{0i}=0$
    automatically.}.

We shall find the $(\vec{\xi},\eta)$ coordinate transformation by
comparing the expression for $K$ in the $(\vec{x},t)$ coordinate
system with $K$ in the $(\vec{\xi},\eta)$ coordinate system.

In the $(\vec{x},t)$ coordinate system, we know that $K$ must be a
linear combination of the ten Killing vectors of Minkowski space:
\begin{equation}
\label{E:general_timelike_Kvector}
    K =\sum_{\alpha=0}^{d} \tau^\alpha T_{(\alpha)} +
       \sum_{\substack{i,j=1}}^{d} \tilde{\rho}^{ij} R_{(ij)}+
       \sum_{i=1}^{d} \beta^i \vec{B}_{(i)}.
\end{equation}
where $\tilde{\rho}_{ij}$ is an antisymmetric matrix,
$T_{(\alpha)}$ are the Killing vectors corresponding to
translations, $R_{(ij)}$ are Killing vectors corresponding to
rotations in the $x_i-x_j$ plane, and $B_{i}$ to rotations in the
$(t-x_i)$ plane (boosts in the $x_i$ direction). In order for $K$
to be time-like, we need that either $\tau_0 \neq 0$ or that
$\beta_i\neq 0$ for some $i$ (or both). In the latter case, this
allows for several simplifications of equation
(\ref{E:general_timelike_Kvector}).

\subsection{The $\vec{\beta}\neq 0$ case}

We may simplify the boost term, $\sum_{i=1}^{d} \beta^i
\vec{B}_{(i)}$ in equation (\ref{E:general_timelike_Kvector}), by
rotating the $\vec{x}$ coordinates \footnote{
    The fact that rotations in the $(\vec{x},t)$ coordinate system do not
    change the stationary character of the metric in the $(\xi,\eta)$
    coordinate system follows from the fact that rotations (and
    similarly translations and boosts) are Killing transformations and
    leave the metric invariant.}:
a linear combination of two boost generators is a boost generator
in another direction, so by choosing this direction to be, say,
the $1$ direction, we may bring (by a spatial rotation in the
cartesian $(\vec{x},t)$ coordinate system) the expression
$\sum_{i=1}^{d} \beta^i \vec{B}_{(i)}$ to the form $\beta
\vec{B}_{(1)}$.

The translation in the direction of the boost may be eliminated by
a shift in the origin of the time coordinate---that is, the axis
of rotation for the boost will be moved. Explicitly $\beta
B_{(x)}+a T_{(x)} = \beta (t+\frac{a}{\beta})\partial_x-\beta
x\partial_t$.

Next consider the time-translation. Adding a time translation to a
boost results in another boost with a shifted space coordinate, so
we may also eliminate the time translation: $\beta B_{(y)}+\tau^0
T_{(0)} = \beta t\partial_y-(\beta y+\tau^0)\partial_t$.

Finally, rotations that rotate the direction of the boost may be
eliminated by changing the boost axis by an appropriate
redefinition of the coordinates. We have thus simplified
expression (\ref{E:general_timelike_Kvector}) to:
\begin{equation}
\label{E:simplified_K}
    K =\sum_{\alpha=2}^{d} \tau^\alpha T_{(\alpha)} +
       \sum_{\substack{i,j=2}}^{d} \tilde{\rho}^{ij} R_{(ij)}+
       \beta \vec{B}_{(1)}.
\end{equation}

In order to find the desired coordinate transformation (equation
(\ref{E:coortans})), we compare the expressions we have for $K$.
On one hand we know from vector transformation properties that:
\begin{equation}
\notag
    \frac{\partial}{\partial \eta} =\frac{\partial t}{\partial \eta} \partial_t
                    +\frac{\partial x}{\partial \eta}\partial_x
                    +\frac{\partial z_i}{\partial \eta}\partial_{z_i}.
\end{equation}
On the other hand, we have from equation (\ref{E:simplified_K})
\begin{equation}
\label{E:set_of_trans_eqns1}
    \partial_{\eta} =\beta x \partial_t
                    +\beta t \partial_x
                    +\left(
                         \tau_i
                        +\sum_{j\neq i>1}^d \tilde{\rho}_{ji}{z}_j
                     \right)
                         \partial_{z_i}.
\end{equation}
Where $x$ is the direction along the boost, and $\vec{z}$ are the
directions orthogonal to it. We get the following set of
equations:
\begin{subequations}
\label{E:set_of_trans_eqns2}
\begin{align}
\label{E:set_of_trans_eqns2a}
    \frac{\partial t}{\partial \eta}&=\beta x \\
\label{E:set_of_trans_eqns2b}
    \frac{\partial x}{\partial \eta}&=\beta t \\
\label{E:set_of_trans_eqns2c}
    \frac{\partial {z}_i}{\partial \eta}&=\tau_i + \tilde{\rho}_{ji} {z}_j
\end{align}
\end{subequations}

The general solution for the first two equations is
\begin{subequations}
\begin{align}
\label{E:gsolution}
    t&=A(\vec{\xi})\sinh(\beta \eta+\Phi(\vec{\xi}))\\
    x&=A(\vec{\xi})\cosh(\beta \eta+\Phi(\vec{\xi}))
\end{align}
\end{subequations}
up to an exchange of the $\cosh$ and $\sinh$ functions (which will
amount to a redefinition of the time and space coordinates.)

We can write the equation for the orthogonal coordinates as a
vector equation:
\begin{equation}
\label{E:equation_for_xperp}
    \dot{\vec{z}}=\vec{\tau} + \tilde{\rho} \vec{z}.
\end{equation}
This is a non homogeneous linear equation. The most general
solution is given by ${\vec{z}}=O(\eta) \vec{h}(\xi) +
\vec{z}_S(\eta)$. Where $O_{ij}(\eta) h_j (\xi)$ (With
$O(\eta)=e^{\tilde{\rho}\eta}$ an orthogonal matrix, and
$\vec{h}(\xi)$ an arbitrary vector) is a solution to the
homogeneous equation, and $z_S(\eta)$ is an arbitrary solution of
the non homogeneous equation. We may solve for $z_S(\eta)$
explicitly, however,  the expression is not necessary for the rest
of our calculations, and it is not very illuminating (note that
$\tilde{\rho}$ is not necessarily invertible.) We emphasize that
$\vec{z}_S(\eta)$ is independent of $\vec{\xi}$, since one can
always treat $\vec{z}$ as a function of $\eta$ only when looking
for a special solution of (\ref{E:equation_for_xperp}).

To simplify matters further, we note that one may arbitrarily
reparametrize the $\vec{\xi}$ coordinates, without changing the
stationary properties of the metric. We shall use this freedom to
define new spatial coordinates: $\zeta_1=A(\vec{\xi})$, and
$\zeta_i=h_i(\vec{\xi})$,
\begin{subequations}
\label{E:gsolution2}
\begin{align}
    t&=\zeta_1 \sinh(\beta \eta+\Phi(\zeta))\\
    x&=\zeta_1 \cosh(\beta \eta+\Phi(\zeta))\\
    {z}_i&=O_{ij} \zeta_j + {z_S}_i(\eta).
\end{align}
\end{subequations}

Now we impose the boundary conditions specified by eq.(\ref{E:bc})
(in Euclidian space). Starting with the first boundary condition
(equation (\ref{E:bc1})), we need that the hypersurface
$\{(\vec{x},0^{+}) | \vec{x} \in V\}$ coincide with the
hypersurface $\{(\vec{\zeta},0) | \vec{\zeta} \in I_{\zeta}\}$. To
verify this we rotate the time coordinates in equations
(\ref{E:gsolution2}) to the imaginary axis, and check that the
hypersurfaces are equal. We get:
\begin{subequations}
\label{E:bc1_for_gsolution1}
\begin{align}
    x&=\zeta_1 \\
    \vec{z}&=\vec{\zeta}_{\bot} + \vec{z}_S(0).
\end{align}
\end{subequations}
Note that $\vec{z}_S(0)$ is a constant vector.

Next, we check the equality of the hypersurfaces at $t=0^-$
(equation (\ref{E:bc2})). We get that in this case:
\begin{subequations}
\label{E:bc2_for_gsolution1}
\begin{align}
    0&=\zeta_1 \sin(\beta \beta_0)\\
    x&=\zeta_1 \cos(\beta \beta_0)\\
    \vec{z}&=O(\imath \beta_0) \vec{\zeta}_{\bot} + \vec{z}_S(\imath \beta_0).
\end{align}
\end{subequations}
This implies that $\beta=\frac{2\pi}{\beta_0}$, and that
\begin{equation}
\label{E:bc1_compared_to_bc2_for_sol1}
    O(\imath \beta_0)
        \vec{\zeta}_{\bot} + \vec{z}_S(\imath \beta_0) =
        \vec{\zeta}_{\bot}+\vec{z}_S(0)
\end{equation}
In order for this equality to hold one needs that the $\zeta$
dependent terms be equal, and that the constant terms be equal.

Starting with the $\zeta$ dependent terms we have: $O(\imath
\beta_0)_{ij} \zeta_j = \zeta_i$. Therefore, $O(\imath
\beta_0)=\delta_{ij}$, implying that $\tilde{\rho}=0$. In this
case $\vec{z}_S(\eta)$ simplifies to
$\vec{z}_S(\eta)=\vec{\tau}\eta$, which according to equation
(\ref{E:bc1_compared_to_bc2_for_sol1}) gives us $\vec{\tau}=0$.

We find that the desired coordinate transformation must satisfy:
\begin{align*}
    t&=\zeta_1 \sinh(\beta \eta)\\
    x&=\zeta_1 \cosh(\beta \eta)\\
    \vec{z}&=\vec{\zeta}_{\bot}.
\end{align*}
This is a polar coordinate transformation in the $(\vec{x},t)$
plane. In order that it be onto, we need that $V$ (assumed to be
connected) is half of space.

\subsection{The $\tau_0\neq 0$ case}
Here, we do not need to simplify equation
(\ref{E:general_timelike_Kvector}), as all the $\beta_i$ vanish.
The equivalent of equations (\ref{E:set_of_trans_eqns2}) are
\begin{align*}
    \frac{\partial t}{\partial \eta}&=\tau_0 \\
    \frac{\partial x}{\partial \eta}&=\tau_i +\sum_{j , i}^{d-1}\tilde{\rho}_{ji}x_j
\end{align*}
Whose solution is:
\begin{align*}
    t&=\tau_0 \eta + A(\vec{\zeta})\\
    z_i&=O_{ij} \zeta_j + z_{S_i}(\eta)
\end{align*}

Imposing the first boundary condition (equation (\ref{E:bc1})), we
get that $A=0$, and that $\vec{z}=\vec{\zeta} +\vec{z}_S(0)$. For
the second condition to hold (equation (\ref{E:bc2})), we need
that $\tau_0=0$ which is ruled out since then $K$ is not timelike.
We note that there is a further possibility: if we identify the
limits $t\to \infty$ and $t \to -\infty$ (meaning, we compactify
the time coordinate on an infinitely large circle), then in the
limit where $\beta_0$ is of the order of the circumference of the
time direction ($\beta \to \infty$), we may have
$\lim\limits_{\eta \to \beta_0} \tau_0 \eta = 0^{-}$. In this
case, we also get $\tilde{\rho}=0$ and $\vec{\tau}=0$ (following
arguments similar to the $\vec{\beta} \neq 0$ case), so that:
\begin{align*}
    t&=\tau_0 \eta \\
    z_i&=\zeta_j,
\end{align*}
which may be reduced to the identity transformation, giving back
Minkowski space. In order for the transformation to be onto, we
obviously need that $V$ be all of space.

What we have shown then, is that if we wish to identify $H_{eff}$
as the Hamiltonian of a free field in a static curved background,
then $V$ must be half of space or all of space, corresponding to a
Rindler coordinate transformation, or no transformation at all.

\section{Implications and summary}

We have discussed two types of observers. One makes measurements
which are restricted to a certain volume $V$ of Minkowski space.
It has been shown elsewhere \cite{HDR,tbp}, that this observer's
measurements will have fluctuations which scale linearly with the
surface area. A second observer has no access to anything external
to $V$. This observer sees a state described by a density matrix
$\rho_{in}$. The averages she will measure are equal to
expectation values of measurements of the first observer who
chooses to restrict his measurements to $V$. This was shown in
section II.

Since fluctuations of measurements for both observers will,
generally, scale linearly with the surface area, we have looked
for the `effective Hamiltonian' of the second observer. It was
shown that if the volume is half of space, there exists a
`natural' Hamiltonian for such an observer: the Rindler
Hamiltonian. Otherwise, we have not found an interpretation of
$H_{eff}$. We did show that it can not be the Hamiltonian of a
free field in a static curved, or geometrically non-trivial,
background.

In the case where the volume is half of space,  the Rindler
observer and the observer restricting his measurements to half of
space are equivalent. This enables us to calculate statistical
averages for the Rindler observer by calculating quantum
expectation values of the Minkowski observer. (This is related to
the fact the left Rindler wedge is the thermofield double of the
right Rindler wedge.) We have used this correspondence to
explicitly calculate the heat capacity of Rindler space radiation,
by calculating the quantum fluctuations of the Rindler Hamiltonian
in the Minkowski vacuum.

We should emphasize, that even though we have no knowledge of its
attributes, the system described by $\rho_{in}$ exhibits area
scaling of fluctuations. Since non-extensivity of fluctuations
exists for any volume, area scaling for a volume which is half of
space is only a special case of area scaling of thermodynamic
quantities. In this sense, one may argue that the area scaling of
Rindler space thermodynamics is a special case of a more general
phenomenon.

Finally, we have shown in \cite{HDR,Implicationstbp} that the area
scaling of fluctuations can lead to a bulk-boundary
correspondence. With our proposed relation between Rindler space
and an observer making measurements in half of Minkowski space,
this may lead to an interesting analogy with a presumed
holographic reduction of Rindler space.

\section{Acknowledgments}

Research supported in part by the Israel Science Foundation under
grant no. 174/00-2 and by the NSF under grant no.PHY-99-07949.
A.~Y. is partially supported by the Kreitman foundation.  R.~B.
thanks the KITP, UC at Santa Barbara, where this work has been
completed. We would like to thank S. de Alwis, and Y. Lederer for
helpful discussions.

\begin{appendix}

\section{Calculation of energy fluctuations in Rindler space.}
\label{A:calculateED2}
We briefly go over the calculations in \cite{tbp}, which give the
energy fluctuations of the Rindler space Hamiltonian in half of Minkowski space.

We first recall the calculation of energy fluctuations in a half
of Minkowski space. We define the energy operator for a general
sub-volume $V$ as
\[
    E^V=\int_V \mathcal{H}(\vec{x}) d^dx,
\]
where $\mathcal{H}(\vec{x})$ is the $T^{00}$ component of the
energy momentum tensor, and we are working in the Schroedinger
picture. For a free field theory, we have
\begin{multline*}
    :E^V:=\frac{1}{4}\frac{1}{(2\pi)^{2d}}
        \int
        \bigg(
            \left(\frac{-\vec{p} \cdot \vec{q}}
                    {\sqrt{\omega_p \omega_q}} -
                    \sqrt{\omega_p \omega_q}\right)
            \left(a_{\vec{p}}a_{\vec{q}}+
                a_{-\vec{p}}a_{-\vec{q}} \right)+
                2\left(\frac{-\vec{p} \cdot \vec{q}}
                    {\sqrt{\omega_p \omega_q}} +
                    \sqrt{\omega_p \omega_q}\right)
            a_{-\vec{p}}a_{\vec{q}}
        \bigg)\\
        \times
        e^{\imath(\vec{p}+\vec{q})\cdot \vec{x}}
        d^dp\,d^dq\,d^dx.
\end{multline*}
We see that $\langle:E_V:\rangle=0$, and proceed to calculate
$\langle:E_V:^2\rangle$. We get:
\begin{equation}
\label{E:HV2}
    \langle:E^V:^2\rangle=\frac{1}{8} \frac{1}{(2\pi)^{2d}}
              \int \int \int_V \int_V
                {\left(\frac{-\vec{p} \cdot \vec{q}}
                    {\sqrt{\omega_p \omega_q}} -
                    \sqrt{\omega_p \omega_q}\right)}^2
                    e^{\imath(\vec{p}+\vec{q})\cdot (\vec{x}-\vec{y})}
                d^dp\,d^dq\,d^dx\,d^dy.
\end{equation}
Since after integration over the momenta we get an integration
over the variable $|\vec{x}-\vec{y}|$, we may rewrite this
integral as
$
    \int \limits_0^\infty F(\xi) D(\xi) d\xi,
$
where for a free field theory,
\begin{align*}
    F(x)&=\frac{1}{8} \frac{1}{(2\pi)^{2d}}
              \int\int
                {\left(\frac{-\vec{p} \cdot \vec{q}}
                    {\sqrt{\omega_p \omega_q}} -
                    \sqrt{\omega_p \omega_q}\right)}^2
                e^{\imath(\vec{p}+\vec{q})\cdot (\vec{x}-\vec{y})}
                d^dp\,d^dq\\
        &=\frac{1}{8} \frac{1}{(2\pi)^{2d}}
           \int
            \left( pq+
                2\vec{p} \cdot \vec{q}+
                \frac{(\vec{p} \cdot {\vec{q}})^2}{pq} \right)
            e^{-\imath(\vec{p}+\vec{q})\cdot \vec{x}}
                d^dp\,d^dq\\
\intertext{and}
    D(\xi)&=\int_V \int_V
            \delta^{(d)}(\xi-|\vec{x}-\vec{y}|)
            d^dx \, d^dy.
\end{align*}

We first consider $F(x)$. We switch to a coordinate system
where:
\[
    \vec{x}=
        \begin{pmatrix}
        x \\
        0 \\
        0 \\
        \vdots \\
        0 \\
        \end{pmatrix}
    ;\,
    \vec{q}=
        \begin{pmatrix}
        q_x \\
        q_{\bot} \\
        0 \\
        \vdots \\
        0 \\
        \end{pmatrix}
        ;\,
    \vec{p}=
        \begin{pmatrix}
        p_x \\
        p_{\bot} \cos \theta_p \\
        p_{\bot_{\bot}} \\
        \vdots \\
        0 \\
        \end{pmatrix}.
\]
In this form, we may do all angular integrations:
\begin{multline*}
    F(x)=\frac{1}{8} \frac{1}{(2\pi)^{2d}}
           \left(
                \frac{\pi^{\frac{d}{2}}}{\Gamma(\frac{d}{2}+1)}(d-1)
           \right)^2\\
           \times
           \int
                \left( pq+
                    2p_xq_x+
                    \frac{p_x^2q_x^2}{pq}+
                    \frac{p_{\bot}^2q_{\bot}^2}{pq}\frac{1}{d-1}
                \right)\\
            \times
                e^{-\imath(p_x+q_x)x}
                p_{\bot}^{d-2}
                q_{\bot}^{d-2}
            dp_{\bot}\,dq_{\bot}
            dp_x\,dq_x.
\end{multline*}
Switching to polar coordinates in the remaining two dimensional system:
\begin{align*}
    p_{\bot}&=p \sin \theta \\
    p_x&=p \cos \theta,
\end{align*}
and observing that integrations over the $p$ and $q$ variables are
independent, we can now evaluate the integral with an exponential
high momentum cutoff:
\begin{align*}
    F(x)&=\frac{(d+1)
                  \Gamma\left(\frac{d+1}{2}\right)
                  \Lambda^{2(d+1)}}
                 {8 \pi^{d+1}
                  (1+(\Lambda x)^2)^{d+3}}
            (d-2(d+2)(\Lambda x)+d(\Lambda x)^4\\
          &=\frac{(d+1)
                  \Gamma\left(\frac{d+1}{2}\right)
                  \Lambda^{2(d+1)}}
                 {8 \pi^{d+1}}
            \nabla^2_{\Lambda x}
            \frac{(\Lambda x)^2-1)}
                 {2(d+2)(1+(\Lambda x)^2)^{d+1}}.
\end{align*}

Next we consider the geometric term for the specific case that $V$
is half of Minkowski space.
\[
    D(\xi)=\int \limits_{-\infty}^{\infty}
             \ldots
             \int \limits_{-\infty}^{\infty}
             \int \limits_0^\infty
             \int \limits_0^\infty
             \delta^{(d)}(\xi-|\vec{x}_1-\vec{x}_2|)
             d^dx_1 \, d^dx_2.
\]
Switching to $\vec{r}_{\pm}=\vec{x}_1 \pm \vec{x}_2$ coordinates,
we may integrate over the transverse $\vec{r}_+$ directions,
yielding the transverse volume $V_{\bot}$ (here transverse means
the directions transverse to the $z$ coordinate.) Therefore:
\[
    D(\xi)=\frac{1}{2} V_{\bot}
             \int \limits_0^\infty
             \int \limits_{-z_+}^{z_+}
             \left[
                \int \limits_{-\infty}^{\infty}
                \ldots
                \int \limits_{-\infty}^{\infty}
                \delta^{(d)}(\xi-r_-)
                d^{d-1}{r_-}_{\bot}
             \right]
             dz_-
             dz_+.
\]

Putting this back in equation (\ref{E:HV2}) and substituting $r_- \to \Lambda r_-$, we have
\begin{equation}
\label{E:Hvsquared}
       \langle:{E^V}^2:\rangle=\frac{1}{2} V_{\bot} \Lambda^{d-1}
             \int \limits_0^\infty
             \int \limits_{-z_+}^{z_+}
             \left[
                \int \limits_{-\infty}^{\infty}
                \ldots
                \int \limits_{-\infty}^{\infty}
                F_d(\frac{r_-}{\Lambda})
                d^{d-1}{r_-}_{\bot}
             \right]
             dz_-
             dz_+.
\end{equation}
Working in a d-dimensional `cylindrical' coordinate system
($d\vec{x}^2=d\rho^2+dz_-^2+\rho^{d-2}d\Omega^2$), we may easily
do all angular integrations. So that
\begin{multline*}
       \langle:{H^V}^2:\rangle\propto
             \int \limits_0^\infty
             \int \limits_{-z_+}^{z_+}
             \int \limits_0^\infty
             \bigg[\frac{\partial}{\partial \rho}
                \left(\rho^{d-2} \frac{\partial}{\partial \rho}
                    \left(\frac{\rho^2+z_-^2-1}
                               {2(d+2)(1+\rho^2+z_-^2)^{d+1}}\right)
                \right)\\
                +\rho^{d-2} \frac{\partial^2}{\partial z_-^2}
                    \left(\frac{\rho^2+z_-^2-1}
                               {2(d+2)(1+\rho^2+z_-^2)^{d+1}}\right)
             \bigg]
             d\rho
             dz_-
             dz_+,
\end{multline*}
where numerical coefficients have been omitted for clarity.  The
first term vanishes upon integration, and the second term may be
calculated explicitly. The final result is:
\[
    \langle:{E^V}^2:\rangle=V_{\bot} \Lambda^{d+1}
              \frac{(d+1) \Gamma^2\left(\frac{d+1}{2}\right)}
                   {2^{d+5} \pi^{1+\frac{d}{2}}
                    \Gamma(2+\frac{d}{2})}.
\]
Hence energy fluctuations are proportional to the surface area
$V_{\bot}$.

As the Rindler Hamiltonian is in essence the `boost' generator, we
proceed to calculate fluctuations of the boost generator in the
`z' direction, $B_{(z)}^V$. Since \cite{Weinbergfieldtheory1}
\[
    B_{(z)}^V=\int \limits_{V} z\mathcal{H}d^dx.
\]

Using the same notation as in the previous calculation, we get
that $\langle:B_{(z)}^V:\rangle=0$, and
\begin{align*}
    \langle:(B_{(z)}^V)^2:\rangle&=\int z_1 z_2 F_d(r_-) d^dx_1\,d^dx_2\\
             &=\frac{1}{2}V_{\bot}
               \int \limits_0^{\infty}
               \int \limits_{-z_+}^{z_+}
               \left[
                    \int \limits_{-\infty}^{\infty}
                    \ldots
                    \int \limits_{-\infty}^{\infty}
                    \frac{1}{4}(z_+^2-z_-^2)
                    F_d(r_-)
                    d^{d-1}{r_-}_{\bot}
               \right]
               dz_-\,dz_+.
\end{align*}
This integral may now be evaluated as before, giving
\[
    \langle:(B_{(z)}^V)^2:\rangle=\frac{1}{\Lambda^2}
                      \frac{1}{d-1}
                      \langle:E^V:^2\rangle.
\]
Here, again, the fluctuations are proportional to the surface
area. The Rindler Hamiltonian is given by $H_R = a B_{(z)}^V$, so
that $\langle :H_R:^2 \rangle = a^2 \langle :
(B_{(z)}^V)^2:\rangle$.

\end{appendix}

\end{document}